# On the bonding environment of phosphorus in purified doped single-walled carbon nanotubes

Georgina Ruiz-Soria[1*], Toma Susi[1], Markus Sauer[1], Kazuhiro Yanagi[2], Thomas Pichler[1], Paola Ayala[1*]

[1] *Faculty of Physics, University of Vienna, Boltzmanngasse 5, A-1090 Vienna, Austria*
[2] *Department of Physics, Tokyo Metropolitan University, 1-1 Minami-Osawa, Hachiouji, Tokyo 192-0397, Japan*

**Abstract**

In this work, phosphorous-doped single-walled carbon nanotubes have been synthesized by the thermal decomposition of trimethylphosphine using a high-vacuum chemical vapor deposition method. Furthermore, a modified density-gradient-ultracentrifugation process has been applied to carefully purify our doped material. The combined use of Raman and X-ray photoelectron spectroscopy allowed us to provide the first insight into the bonding environment of P incorporated into the carbon lattice, avoiding competing signals arising from synthesis byproducts. This study represents the first step toward the proper identification of the bonding configuration of P atoms when direct substitution takes place.

## 1. Introduction

Fine-tuning the electronic properties of single-walled carbon nanotubes (SWCNTs) via different functionalization paths is highly desirable for applications. The direct replacement of C atoms by heteroatoms is an ideal method, but also the one presenting the largest difficulties [1, 2]. Experimentally, substitutional nitrogen doping has already been extensively explored and the bonding configurations that N atoms can have in the graphitic network of a nanotube have been studied by various methods [3, 4]. In a more limited number of publications, boron has also been reported as a feasible substitutional element [5, 6, 7, 8]. Another dopant candidate is phosphorus. Calculations have indicated that P can incorporate in the nanotube wall, taking the place of a C atom through direct substitution. It has been suggested that P incorporation allows the modulation of the transport properties of CPx-SWNTs [2, 9]. Although the experimental incorporation of P in carbon materials such as diamond and fullerenes was reported more than a decade ago [10, 11], the theoretical prediction of P-doped single-walled carbon nanotubes (CPx-SWNTs) has been followed only much later by experimental results using aerosol-assisted chemical vapor deposition (CVD) [12] and arc discharge [13]. Similarly to other doped SWCNTs, even in the optimal conditions, the properties of produced CPx-SWNTs depend on the specific P precursor and the catalytic method used for synthesis.

In this context, it has been reported that P and N working as co-dopants can favor the oxygen reduction in proton exchange membrane fuel cells [14]. Enhanced charge carrier spin scattering compared to pristine SWCNTs has been reported from nuclear magnetic resonance measurements [13]. Nevertheless, the field is still in its initial stages, and an understanding of the material's properties as well as the optimization of the synthesis processes is needed before CPx-SWNTs can find their way into applica-

*Corresponding authors. Fax:+43-1427751375.
E-mail addresses:
georgina.ruiz@univie.ac.at (G.Ruiz-Soria);
paola.ayala@univie.ac.at (P.Ayala)

tions. It is necessary to understand the incorporation of P atoms in the nanotube to analyze whether the overall morphology of the tubes is affected by the purification process, i.e. the tube diameter and the sample diameter distribution

To address this matter, we have used for the first time a non-diluted trimethylphosphine feedstock for the synthesis of CPx-SWNTs using a high-vacuum based CVD method. We have noticed that for the specific case of CPx-SWNTs, the yield of nanotubes is reduced in comparison to previously reported CVD methods. However, the possibility to use a non-diluted precursor provides a great advantage towards understanding the bonding environments in which P bonds to the carbon atoms in the graphitic lattice as well as to the catalytic byproducts. Only indirect techniques, namely Raman spectroscopy and transport measurements, had so far been used to infer the incorporation of P in the nanotubes [9, 13]. To the best of our knowledge, none of the reports on CPx-SWNTs have been able to provide direct evidence for the incorporation of P atoms in substitutional configuration in a single-walled nanotube sample nor an estimation of the bulk percentage of P heteroatoms.

In this context, the purification of CPx-SWNTs we report here has allowed analytically disentangling the role of P in the nanotubes exclusively. We have benefited from the use of a purified material to apply the full capability of X-ray photoelectron spectroscopy (XPS) to understand the bonding environments of P in a SWCNT doped sample.

2. Experimental

A high-vacuum (HV) CVD, was used for the thermal decompostion of trimethylphosphine (PMe$_3$) in the presence of an iron-based supported catalyst. This is the first work to report the use of this precursor and HV-CVD system to synthesize CPx-SWNTs nanotubes. This method has been used previously for the synthesis of pristine and doped SWCNTs as described in detail elsewhere [15, 16, 17]. The optimal growth temperature for CPx-SWNTs was found to be ∼900 °C. The samples were purified using a modified density gradient ultracentrifugation (DGU) method [18, 19, 20] using a mild acid solution of HCl in water (30 wt.%). The metallic catalytic particles were eliminated by tip sonication in a 1% sodium deoxycholate (DOC) aqueous solution for 4 h and later ultracentrifugation for 30 min at ∼150000 g. From the centrifugation tube, only ∼ 70% of the supernatant was collected, since the upper ∼ 10% is typically constituted by remaining amorphous carbon and the bottom ∼ 20% of metallic particles. To clean the surfactant from the nanotubes, the solution was filtrated through PTFE membrane filters (Milipore, pore size 0.2 $\mu$m) and rinsed thoroughly with methanol, water and toluene. The overall morphology of the samples was inspected by scanning electron microscopy (SEM; Zeiss Supra 55 VP, 1.00 kV gun acceleration voltage) and transmission electron microscopy (TEM; FEI Tecnai F30, operating at 100 keV). A JY-Horiba HR800 Raman spectrometer with a liquid nitrogen cooled CCD detector was used for multifrequency measurements using 488, 568 and 632 nm excitation wavelengths. All data were collected in ambient conditions. XPS was used to inspect the elemental composition of the samples, with the spectra recorded using a Scienta spectrometer equipped with a MX650 monochromatic X-ray source.

3. Results and Discussion

The optimal growth temperature was identified to be ∼900 °C. However, the growth temperature is practically limited to a T 50°C window. Out of this range the growth of nanotubes was practically voided. Based on this fact, we focused on the quality of the experiments with steps of T 10°C. The yield was indeed slightly compromised away from 900°C but no significant morphology changes were identified. The first remarkable result regards the possibility to obtain a buckypaper of purified doped material. The limited number of works related to the purification of either in-situ or post synthesis doped tubes had previously highlighted this as a problem [20, 21]. SEM and TEM micrographs, as well as the schematics of the purification procedure are shown in Fig. 1.

The low magnification image displays highly compact bundles of CPx-SWNTs with no visible impurities. To provide a rough estimation of the distri-bution of diameters in the raw and purified samples, a multifrequency resonance Raman spectroscopy inspection of the radial breathing mode (RBM) region is shown in (Fig. 2a). It is well understood that the Raman response in this region, $\omega_{RBM}$, is inversely proportional to the diameter of the tubes $d_t$. We have made use of the environmental parameters and force constant reported in Ref. 22) for our estimation. From our measurements with the three laser lines, the highest diameter populations were found in the range between $\sim$0.85 and $\sim$1.4 nm (dotted vertical lines). The spectra in Fig. 2 are normalized to the maximum peak intensity for ease of comparison. However, comparing the spectra of the raw material with the purified one with the intensity normalized to the laser power (not shown here) reveals a particularly pronounced elimination of tubes with diameters larger than $\sim$1.2 nm, which can be attributed to the centrifugation step. Small diameter tubes are not abundant in the material as-grown and the purification reduces their population further, most likely due to their lower stability during ultrasonication. Thus the mean diameter in the sample remains almost unchanged upon purification. To get a Raman estimate of the defect concentration in our samples, we use the D band to G band intensity ratio ($I_D/I_G$). Fig. 2b shows the D and G band responses for the raw and the purified material. The $I_D/I_G$ intensity ratio confirms the high quality of the material with clearly discernible transverse and longitudinal optical phonon signals, and it also indicates that the purification process results in significantly cleaner tubes (see Table 1).

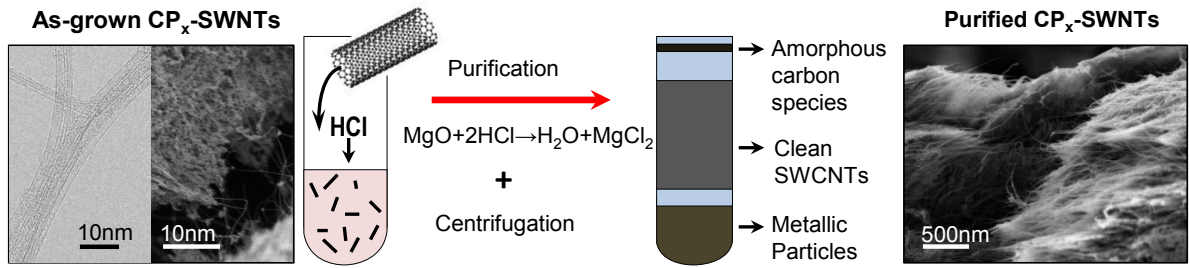

Figure 1: Intermediate magnification TEM image and SEM micrograph of the sample as grown (left), schematics of the DGU purification process (middle), and SEM micrograph of the material after purification (right).

| Laser wavelength (nm) | As-grown $I_D/I_G$ | Purified $I_D/I_G$ | Change (%) |
|---|---|---|---|
| 488 | 0.2 | 0.07 | −65 |
| 568 | 0.5 | 0.27 | −46 |
| 633 | 0.6 | 0.48 | −20 |

Table 1: Changes of the Raman intensity ratio of the D band to the G band upon purification according to the excitation wavelength.

We have examined carefully the 2D line depicted in Fig. 2c. It has been previously reported that this line should exhibit a clear splitting due to an associated double phonon scattering process induced by the presence of dopants [23]. Similar to those studies, the spectra corresponding to our non-purified CPx-SWNTs exhibit a much larger full width at half maximum (FWHM) compared to the 2D response of the purified material, which can in principle be deconvoluted with two different contributions. However, after the sample has been purified, no splitting of the 2D band is observed. Therefore, it remains uncertain at what doping level the 2D signature could be used to quantify the incorporation of dopants.

In order to verify the purity of our samples, to analyze their elemental composition and to confirm the presence of phosphorus, we performed XPS measurements. It has been reported that phosphorus dopants can get strongly oxidized in ambient conditions [13], we therefore recorded spectra before and after an-nealing the sample in vacuum to remove the adsorbed oxygen. Survey spectra and additional measurements on the energy range corresponding to the Fe $2p$ photoemission response of the purified sample, allowed us to confirm that all rests of catalyst were removed by the purification process (iron in particular), which allow us to probe the incorporation of P in from a different perspective compared to previous XPS studies on CPx-MWNTs [24, 25].

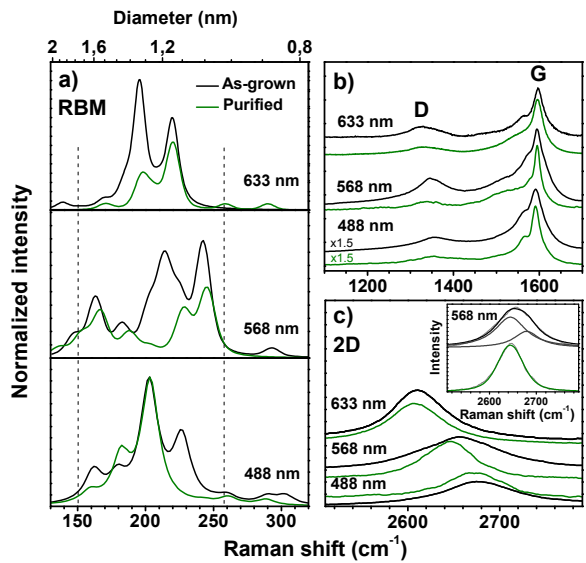

Figure 2: Multifrequency Raman spectra comparing the response of the samples before (black) and after (green) purification. (a) RBM region. (b) D and G band region. (c) 2D band region. The inset shows a close-up to the 2D main peak (568 nm).

In the C $1s$ response for the sample as-grown (not shown here), the main feature present in the spectra is a single intense component located at 284.4 eV, which is in good agreement to the signal corresponding to carbon atoms bound in the graphitic SWCNT lattice [26]. A clean sample of SWCNTs exhibits established footprints in the C $1s$ line that serve as a first prerequisite for the identification of dopants [1]. However, once the samples are purified a broadening of the C $1s$ line is observed. The C $1s$ of the sample after purification, as well as the signal after annealing treatment are shown in the inset of Fig. 3.

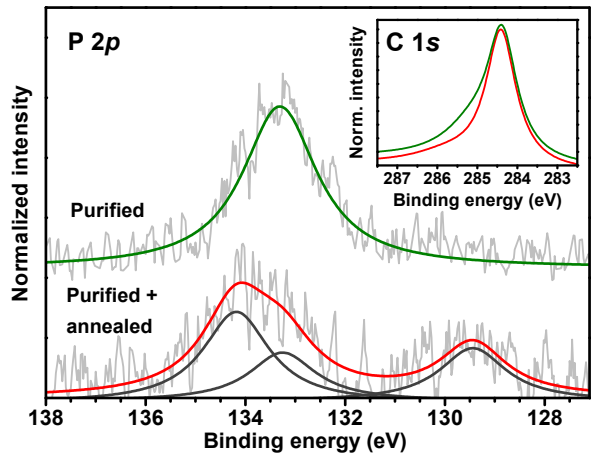

Figure 3: Main panel: P $2p$ core level XPS spectra of the samples after purification and after additional annealing at 600 °C. Inset: the C $1s$ response of the two samples.

This broadening can be partly attributed to oxygen adsorption [27, 28], and it is clearly observed that the FWHM becomes narrower after annealing. Also, the C $1s$ signal exhibited a shift of 0.3 eV compared to the as-grown carbon nanotubes [26]. To try further understanding the C $1s$ lineshape we also need to focus on the phosphorous response. The presence of carbonaceous species associated to phosphorus-containing compounds had been reported as products of the synthesis of multi- and single-walled phosphorus-doped tubes [12, 24, 25], but a direct wall-dopant identification was never brought to light. In our samples, the P content was found to reach 0.3 at.% in the purified sample, and a subsequent *in-situ* annealing step at 600 °C, the phosphorus content was measured as ∼ 0.26 at.%. Thus while the P content did not change significantly, the C $1s$ exhibits a much narrower lineshape (see the inset of Fig. 3).

Considering the spectrometer resolution, the P $2p$ response of the purified sample shown in Fig. 3 can be fitted by a single peak located at 133.3 eV. Note that previous work on C films containing P have [29, 30] attributed this binding energy value to P-O bonds. However, this does not necessarily mean that the P atoms in our samples are bonded to O atoms. Af-ter annealing, this component is still present, but another one at ∼134.2 eV appears. An additional peak at a lower binding energy also becomes apparent, which is not surprising considering the known reactivity of phosphorus. However, this signal at 129.3 eV reported for carbon phosphide films [29] also cannot necessarily be associated to C-P bonds because of the different system in question. However, considering the presence of C, P and O in the elemental composition of the samples even after purification and annealing, we conclude that the signal at 133.3 eV is likely indicative of the incorporation of P substitutionally in the SWCNT's wall. As the only previous work using XPS to probe P doping in nanotubes is related to unpurified multiwalled material [24], this represents the first direct evidence for the incorporation of P into the SWCNT's lattice with this method.

4. Conclusions

We have reported on the first viable synthesis of phosphorus-doped single-walled carbon nanotubes on supported catalysts using high vacuum CVD using a vapor of non-diluted $PMe_3$. As a further advance over most published works on doped SWCNTs, our material was carefully purified to remove byproducts and all catalyst particles. This ensures that our detailed spectroscopic characterization using multifrequency Raman and x-ray photoelectron spectroscopy is probing only the desired SWCNT material. As a result, we are able to unambiguously demonstrate the presence of P in SWCNT samples for the first time. Our P $2p$ photoemission findings suggest the presence of at least three different bonding environments that involve the presence of phosphorus.

**Acknowledgements** This work has been supported by the Austrian Science Fund through project FWF P21333-N20. T.S. has been supported by an FWF grant M 1497-N19, by the Finnish Cultural Foundation and the Walter Ahlström Foundation.